\documentclass[aip,graphicx]{revtex4-1}
\usepackage[T1]{fontenc}
\usepackage{dcolumn}
\usepackage{amsfonts, amssymb, amsmath, bm, graphicx}
\usepackage{etoolbox}

\def\t#1{{\mathrm{#1}}}

\begin{document}


\title{Exchange-mediated directional coupling of short-wavelength spin waves} 



\author{Zhen Zhu}
\thanks{These authors contributed equally to this work.}
\affiliation{School of Physics, Hubei Key Laboratory of Gravitation and Quantum Physics, Institute for Quantum Science and Engineering, Huazhong University of Science and Technology, Wuhan, China}

\author{Roman Verba}
\thanks{These authors contributed equally to this work.}
\affiliation{V. G. Baryakhtar Institute of Magnetism of the NAS of Ukraine, Kyiv, Ukraine}

\author{Philipp Pirro}
\affiliation{Fachbereich Physik and Landesforschungszentrum OPTIMAS, Rheinland-Pfälzische Technische Universität Kaiserslautern-Landau, Kaiserslautern, Germany}

\author{Andrii V. Chumak}
\affiliation{Faculty of Physics, University of Vienna, Vienna, Austria}

\author{Qi Wang}
 \email[Author to whom correspondence should be addressed:]{\ williamqiwang@hust.edu.cn}
\affiliation{School of Physics, Hubei Key Laboratory of Gravitation and Quantum Physics, Institute for Quantum Science and Engineering, Huazhong University of Science and Technology, Wuhan, China}


\date{\today}

\begin{abstract}
Miniaturizing magnonic circuits and increasing their processing speed require spin waves with progressively shorter wavelengths. However, conventional directional couplers rely on the dynamic dipolar interaction between spatially separated waveguides, which rapidly weakens in the exchange-dominated regime. As a result, their coupling length increases as the spin-wave wavelength decreases, limiting further device miniaturization. Here, we introduce an exchange-mediated directional coupler by replacing the nonmagnetic gap between two waveguides with a magnetically modified YIG spacer. The spacer provides a continuous exchange pathway and preserves the splitting between the symmetric and antisymmetric modes at large wavenumbers. We develop an analytical model to describe this mechanism and verify it using micromagnetic simulations. Based on this concept, we design a straight coupler without curved access waveguides, with a footprint of only 400 nm $\times$ 30 nm. For a spin-wave wavelength of approximately 100 nm, the device transfers approximately 98\% of the normalized output power to the target waveguide. Its performance remains nearly unchanged even when the damping of the spacer is strongly increased. These results overcome the short-wavelength limitation of dipolar couplers and provide a route toward highly integrated magnonic circuits for information processing.

\end{abstract}


\maketitle 


Magnonics exploits spin waves---the collective excitations of ordered magnetic moments—and their quanta, magnons, to transmit and process information \cite{chumak2022advances,dieny2020opportunities,flebus20242024,han2024magnonics,pirro2021advances,wang2024nanoscale}. In low-damping magnetic insulators such as yttrium iron garnet (YIG), spin waves can propagate over long distances\cite{liu2018long,qin2022low}, while possessing wavelengths far below those of electromagnetic waves at the same frequency \cite{wintz2016magnetic,che2020efficient,van2016tunable,han2019mutual,urazhdin2014nanomagnonic,pirro2014spin,demidov2016excitation}. This combination makes spin waves attractive for compact wave-based circuits in which information is encoded in amplitude, phase, frequency, or propagation path\cite{kruglyak2010magnonics,szulc2020spin,dobrovolskiy2019spin,lisiecki2019reprogrammability,golkebiewski2022self}. Directional couplers are fundamental building blocks in integrated wave systems and have been widely used in microwave and quantum photonic circuits for signal splitting, routing, interference, and mode manipulation\cite{Moody_2022,1617827}. They also play an important role in integrated magnonics, where coherent spin-wave transfer between neighboring waveguides enables compact signal routing and information processing\cite{szulc2025magnetic,wang2018reconfigurable,sadovnikov2015directional}. 
In a conventional spin-wave directional coupler, the dynamic stray fields of the two waveguides hybridize their eigenmodes into symmetric and antisymmetric collective modes \cite{wang2018reconfigurable,sadovnikov2015directional,ge2025deeply,zhao2022reconfigurable,wang2021stimulated}. Because these modes have different wavenumbers at a fixed frequency, their interference produces a periodic transfer of spin-wave power between the waveguides. This principle enables compact signal routing and reconfigurable magnonic logic\cite{wang2020magnonic}, but the coupling mechanism is intrinsically dipolar.

The reliance on dipolar interaction creates a fundamental scaling problem. Device miniaturization requires shorter-wavelength spin waves and, therefore, operation at larger wavenumbers. In this limit, the exchange interaction increasingly dominates the spin-wave dispersion, while the dynamic dipolar fields responsible for coupling spatially separated waveguides become weak\cite{kalinikos1986theory}. The splitting between the symmetric and antisymmetric branches consequently collapses, the coupling length grows, and complete power transfer becomes impossible within a deeply nanoscale device\cite{wang2018reconfigurable}. Thus, reducing the wavelength alone does not automatically shrink a dipolar directional coupler because the dipolar coupling mechanism disappears upon entering the exchange-magnon regime. Miniaturization is not an end in itself but is also important for reducing the energy cost of information processing\cite{wang2024nanoscale}. In addition, the transition from dipolar to exchange spin waves is accompanied by increases in spin-wave frequency and group velocity, thereby enabling higher processing speeds.

Here, we overcome this limitation by replacing the nonmagnetic gap between the waveguides with a magnetic material. The magnetic spacer provides an exchange pathway across the coupling region, converting the dominant interwaveguide interaction from dipolar to exchange coupling. Unlike the dipolar contribution, the exchange-mediated splitting remains finite at large wavenumbers and can therefore sustain efficient directional coupling for short-wavelength magnons. The concept is compatible with a laterally patterned, planar geometry. The possible implementations are partial laser modification or focused-ion-beam exposure of a continuous YIG film, which can locally reduce its magnetic parameters without physically separating the waveguides \cite{florio2026programmableintegratedmagnonicmeshes,kiechle2023spin,naunheimer2026establishing}.

We formulate an analytical theory for the collective modes of two magnonic waveguides connected by a dissimilar magnetic spacer and compare it with micromagnetic simulations. The theory identifies how the spacer controls the symmetric and antisymmetric mode shifts and clarifies the conditions under which a finite short-wavelength splitting is retained.  We then demonstrate a straight exchange-mediated coupler with a footprint of only 400 nm × 30 nm and approximately 98\% power transfer at 7 GHz. The proposed mechanism directly addresses the short-wavelength breakdown of dipolar couplers and provides a scalable building block for nanoscale magnonic circuits.

To illustrate the scaling limitation of conventional dipolar coupling and the advantage of the proposed exchange-mediated scheme, we first compare the collective-mode dispersions of two coupled YIG waveguides separated by either a nonmagnetic or magnetic spacer. Each YIG waveguide is 50 nm wide\cite{heinz2020propagation} and 10 nm thick, and the spacer between the waveguides is 10 nm wide. The micromagnetic simulations were performed using MuMax3\cite{vansteenkiste2014design}, with further details provided in the Methods section. The waveguides were assigned typical material parameters of YIG\cite{chumak2017magnonic}: a saturation magnetization of $M_{\mathrm{s}} = 140\ \mathrm{kA}/\mathrm{m}$, an exchange stiffness of $A_{\mathrm{ex}} = 3.6\ \mathrm{pJ}/\mathrm{m}$, and a Gilbert damping coefficient of $\alpha = 2 \times 10^{-4}$. Importantly, no external magnetic field was applied. Owing to the shape anisotropy of the elongated waveguides, the equilibrium magnetization remains aligned along the longitudinal waveguide direction, providing a stable remanent magnetic state\cite{demidov/acs.nanolett.3c02725,demidov/sciadv.adx2018}. The spin waves propagate parallel to the equilibrium magnetization and therefore correspond to the backward-volume spin-wave geometry\cite{Serga_2010}.

At a fixed frequency, the coupling strength can be quantified by the wavenumber difference between the symmetric and antisymmetric collective modes, $\Delta k = |k_{\mathrm{s}} - k_{\mathrm{as}}|$. The interference between these two modes produces a periodic transfer of spin-wave power between the waveguides. The length required for complete power transfer is given by $L_{\mathrm{c}} = \pi / \Delta k$. Therefore, a larger separation between the symmetric and antisymmetric modes directly corresponds to a shorter coupling length and a more compact directional coupler.

\begin{figure}[htb]
\centering
\includegraphics[width=0.8\textwidth]{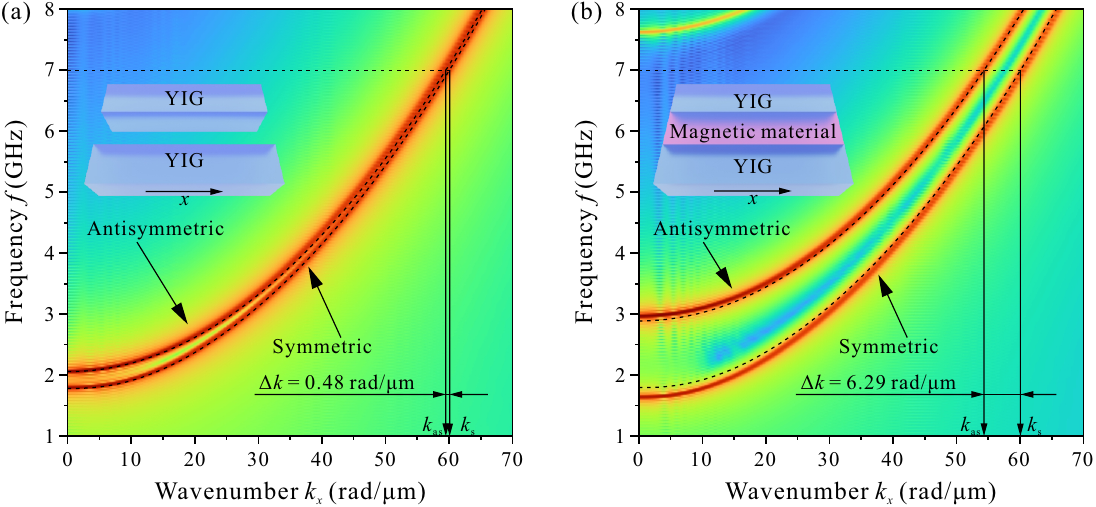}
\caption{\label{fig:epsart}\textbf{Dispersion characteristics of dipole-coupled and exchange-mediated spin-wave directional couplers.} Micromagnetically simulated spin-wave dispersion spectra of two YIG waveguides separated by (a) an air gap and (b) a magnetically modified YIG spacer. The color maps represent the spectral intensity obtained from micromagnetic simulations, while the black dashed curves show the theoretically calculated symmetric and antisymmetric dispersion branches. For the magnetic spacer in (b), the saturation magnetization $M_{s}$ and exchange stiffness $A_\mathrm{ex}$ are both set to 30\% of their respective values in pristine YIG. It is worth noting that the higher-order modes can also be observed in (b). At 7 GHz, the wavenumber difference $\Delta k = |k_{s} - k_{as}|$ increases from $0.48\ \mathrm{rad}/\mu\mathrm{m}$ for the air-gap structure to $6.29\ \mathrm{rad}/\mu\mathrm{m}$ for the magnetic-spacer structure. Accordingly, the coupling length $L_{c} = \pi/\Delta k$ decreases from approximately 6.5 $\mu$m to 0.50 $\mu$m. The insets schematically illustrate the corresponding waveguide structures.}
\end{figure}

Figure~\ref{fig:epsart}(a) shows the micromagnetically simulated dispersion spectrum of two YIG waveguides separated by an air gap, representing a conventional dipole-coupled directional coupler. In the low-wavenumber, dipole-dominated regime, a pronounced separation between the symmetric and antisymmetric branches is observed, indicating strong hybridization mediated by the dynamic stray fields of the two waveguides. As the wavenumber increases, however, the spin waves progressively enter the exchange-dominated regime, while the dynamic stray fields responsible for coupling the spatially separated waveguides become increasingly localized. Consequently, the splitting between the symmetric and antisymmetric modes decreases substantially.

This limitation is particularly evident at $f = 7$ GHz, as indicated by the horizontal dotted line in Fig.~\ref{fig:epsart}(a). At this frequency, the two modes have wavenumbers of approximately $60\ \mathrm{rad}/\mu\mathrm{m}$, corresponding to a spin-wave wavelength of approximately 100 nm. Nevertheless, their wavenumber difference is only $\Delta k = 0.48\ \mathrm{rad}/\mu\mathrm{m}$, which yields a coupling length of $L_{\mathrm{c}} = \pi / \Delta k \approx 6.5\ \mu\mathrm{m}$. The coupling length is thus more than 50 times the spin-wave wavelength and more than 100 times the width of an individual waveguide. This pronounced mismatch demonstrates that reducing the spin-wave wavelength and the transverse waveguide dimensions does not result in a correspondingly short dipole-coupled device. Although both the information carrier and the waveguides have reached the nanoscale, several micrometres of propagation are still required to achieve complete power transfer\cite{wang2018reconfigurable,sadovnikov2015directional}.

To overcome this limitation, we replace the dipolar coupling between the two waveguides with exchange-mediated coupling by introducing a magnetic spacer in place of the 10-nm-wide air gap, as illustrated in Fig.~\ref{fig:epsart}(b). As a representative example, we consider a spacer with the same thickness as the waveguides and set its saturation magnetization \(M_\mathrm{s}\) and exchange stiffness \(A_{\mathrm{ex}}\) to 30\% of the pristine-YIG values. The magnetic order of the spacer is preserved, thereby establishing a continuous exchange pathway between the two waveguides. Such a magnetically modified region could potentially be realized in a continuous YIG film using local laser irradiation or focused-ion-beam exposure\cite{vogel2015optically,florio2026programmableintegratedmagnonicmeshes,kiechle2023spin,greil2025effect,riddiford2025two,naunheimer2026establishing,levati2025three,ruane2018controlling}.

As shown in Fig.~\ref{fig:epsart}(b), introducing the magnetic spacer markedly enhances the separation between the symmetric and antisymmetric modes throughout the investigated frequency and wavenumber range. The enhancement becomes particularly pronounced at large wavenumbers, where dipolar coupling across an air gap is strongly suppressed but exchange-mediated coupling through the magnetic spacer remains effective. At 7 GHz, the wavenumber difference increases to $\Delta k = 6.29\ \mathrm{rad}/\mu\mathrm{m}$, which is more than 13 times the value obtained for the air-gap structure. Accordingly, the coupling length decreases to $L_{\mathrm{c}} = \pi / \Delta k \approx 0.50\ \mu\mathrm{m}$. The magnetic spacer therefore reduces the coupling length by more than one order of magnitude, from approximately 6.5 $\mu$m to 0.50 $\mu$m, while operating with spin waves having wavelengths of approximately 100 nm. These results demonstrate that a continuous magnetic pathway preserves strong mode hybridization in the exchange-dominated regime and enables complete spin-wave power transfer over a submicrometre propagation distance.

The micromagnetic results in Fig.~\ref{fig:epsart} demonstrate that replacing the air gap with a magnetic spacer substantially enhances the mode splitting and reduces the coupling length in the exchange-dominated regime. To clarify the physical origin of this enhancement and determine how the properties of the spacer control the collective modes, we develop an analytical model for two identical magnetic waveguides connected by a dissimilar magnetic material.

We consider two identical waveguides of width $w$, saturation magnetization $M_{\mathrm{s},1}$, and exchange stiffness $A_{\mathrm{ex},1}$, separated by a magnetic spacer of width $d$, saturation magnetization $M_{\mathrm{s},2}$, and exchange stiffness $A_{\mathrm{ex},2}$. The $x$ axis is defined along the waveguides and the $y$ axis along their transverse direction. The equilibrium magnetization is assumed to be uniform and parallel to the $x$ axis in all three regions. We further assume that the dynamic magnetization is uniform across the film thickness and neglect crystalline magnetic anisotropy.

A rigorous analytical solution of the complete dipole-exchange problem is not readily available in closed form. Therefore, we first neglect the dipolar interaction, thus assuming spin waves in the deeply exchange-dominated regime, characterized by circular magnetization precession. The linearized Landau-Lifshitz equation for the dynamic magnetization  $m\left(y\right)$ can then be written as
\begin{equation}
	\left[\omega_0 + \omega_{M,i} \lambda_i^2 \left(k_x^2 - \frac{\partial^2}{\partial y^2}\right) \right] m(y) = \omega m(y) \,,
\end{equation}
where $i=1$ denotes the pristine YIG waveguides and $i=2$ the magnetic spacer,
\begin{equation}
    \omega_{M,i} = \gamma \mu_0 M_{\mathrm{s},i},\ \lambda_i = \sqrt{2 A_{\mathrm{ex},i}/(\mu_0 M_{\mathrm{s},i}^2)}
\end{equation}
$k_{x}$ is the spin-wave wavenumber along the waveguides, and $\omega_{0}$ is the ferromagnetic resonance frequency (which, in the pure exchange approximation, i.e. neglecting magnetostatic interaction at all, is the same in the waveguides and spacer). Finally, $\omega$ is the collective spin-wave mode frequency, which we are searching for.

Assuming free boundary conditions at the two outer edges, $\partial m/\partial y = 0$, the dynamic magnetization profiles in the upper waveguide, magnetic spacer, and lower waveguide can be expressed as
\begin{equation}
	\begin{split}
		m(y) &= C_1 \cos[\kappa_1 (y-y_1)] \,\quad\t{(upper}\,\t{waveguide)},\\ 
		m(y) &= C_2 \cos[\kappa_2 y] + D_2 \sin[\kappa_2 y] \,\quad\t{(spacer)},\\ 
		m(y) &= C_3 \cos[\kappa_1 (y+y_1)] \,\quad\t{(lower}\,\t{waveguide)} \,, 
	\end{split}	
\end{equation}
where $\pm y_{1} = \pm(w+d/2)$ are positions of the outer edges of the waveguides and $y = 0$ is assigned to the spacer center. The transverse wavenumber in each material is given by $\kappa_i^2 = - k_x^2 + (\omega-\omega_0)/(\omega_{M,i}\lambda_i^2)$. At each waveguide-spacer interface, the dynamic magnetization is continuous, while conservation of the exchange torque requires
\begin{equation}
		A_{\t{ex},i} \frac{\partial m_i}{\partial y} = A_{\t{ex},j} \frac{\partial m_j}{\partial y} \,,
\end{equation}
Equivalently, this condition can be written as
\begin{equation}
	\omega_{M,i}^2 \lambda_i^2 \frac{\partial m_i}{\partial y} = \omega_{M,j}^2 \lambda_j^2 \frac{\partial m_j}{\partial y} \ .
\end{equation}

Because the structure is mirror symmetric with respect to the center of the spacer, its eigenmodes are symmetric and antisymmetric collective modes. For the symmetric mode, $D_{2}=0$, and the dispersion is determined by
\begin{equation}\label{e:sym-implicit}
	\omega_{M,1}^2 \lambda_1^2 \kappa_1 \tan[\kappa_1 w] + \omega_{M,2}^2 \lambda_2^2 \kappa_2 \tan\left[\frac{\kappa_2 d}2\right] = 0 \ .
\end{equation}

For sufficiently narrow waveguides and spacers, the symmetric mode is nearly uniform across the structure, such that $\lvert \kappa_1 w \rvert \ll 1$ and $\lvert \kappa_2 d/2 \rvert \ll 1$. Under this approximation, its dispersion can be written as $\omega_\t{\text{S}} \approx \omega_\t{i} + \Delta\omega_\t{\text{S}}$,
where 
\begin{equation}
	\omega_\t{i} = \omega_0 + \omega_{M,1} \lambda_1^2 k_x^2
\end{equation}
is the spin-wave dispersion in an isolated waveguide, and
\begin{equation}\label{e:S-appr}
	\Delta\omega_\t{\text{S}} = \omega_{M,2} k_x^2 d \frac{\omega_{M,2} \lambda_2^2 - \omega_{M,1}\lambda_1^2}{2\omega_{M,1}w + \omega_{M,2}d}
\end{equation}
is the dispersion shift due to the exchange coupling via the spacer.

For the antisymmetric mode, $C_{2}=0$, and the corresponding implicit dispersion relation is
\begin{equation}\label{e:as-implicit}
	\omega_{M,1}^2 \lambda_1^2 \kappa_1 \tan[\kappa_1 w] = \omega_{M,2}^2 \lambda_2^2 \kappa_2 \cot\left[\frac{\kappa_2 d}2\right] \ .
\end{equation}
For a sufficiently narrow spacer, $\cot\left(\kappa_2 d/2\right) \simeq 2/(\kappa_2 d)$, yielding $\omega_\t{\text{AS}} = \omega_\t{i} + \Delta\omega_\t{\text{AS}}$, with
\begin{equation}
    \quad \Delta\omega_\t{\text{AS}} = \omega_{M,1} \lambda_1^2 \left(\frac{\xi}{w} \right)^2.
\end{equation}
Here, $\xi$ is determined by the implicit equation
\begin{equation}\label{e:xi-def}
	\xi \tan \xi = \mathcal{C} = \frac{2\omega_{M,2}^2 \lambda_2^2 w} {\omega_{M,1}^2 \lambda_1^2 d} \ .
\end{equation}

The key result is that the exchange-induced shift of the antisymmetric mode, $\Delta \omega_{\mathrm{AS}}$, is independent of the longitudinal wavenumber $k_{x}$. Physically, the antisymmetric mode possesses a strong transverse variation of the dynamic magnetization, particularly within the narrow spacer, and therefore carries an additional exchange-energy cost. By contrast, the nearly uniform symmetric mode does not acquire such an energy penalty when $M_{\mathrm{s}}$ and $A_{\mathrm{ex}}$ are reduced proportionally. A finite separation between the symmetric and antisymmetric modes therefore remains in the large-$k_x$ regime. This behavior is fundamentally different from dipolar coupling across a nonmagnetic gap, which rapidly weakens as the wavelength decreases.

An interesting feature occurs when the saturation magnetization and exchange stiffness of the spacer are reduced by the same factor $\beta$,
\begin{equation}
	M_{\text{s},2} = \beta M_{\text{s},1}, A_{\text{ex},2} = \beta A_{\text{ex},1}.
	\label{rate}
\end{equation}
Because $\omega_{M} \lambda^{2} \propto A_{\mathrm{ex}} / M_{\mathrm{s}}$, Eq.~(\ref{rate}) gives $\omega_{M,1} \lambda^{2}_1 = \omega_{M,2} \lambda^{2}_2$ and hence the shift of symmetric collective mode $\Delta \omega_{\mathrm{S}}=0$. Thus, the shifts of both the symmetric and antisymmetric modes become independent of spin-wave wavenumber, i.e. their splitting (in frequency units) remains the same and large in the whole wavenumber range. The wavenumber separation of equal-frequency collective waves slowly decreases with $k_x$, since $\Delta k = (\omega_\t{AS}-\omega_\t{S})/(\partial \omega/\partial k_x) \sim 1/k_x$. This feature is observed in Fig.~\ref{fig:epsart}(b), where both $M_{\mathrm{s},2}$ and $A_{\mathrm{ex},2}$ are set to 30\% of their pristine-YIG values.

The proportional scaling in Eq.~(\ref{rate}) is a convenient choice rather than a necessary condition for exchange-mediated coupling. More generally, we may define two independent scaling factors, \(\beta_{M}=M_{\mathrm{s},2}/M_{\mathrm{s},1}\) and \(\beta_{A}=A_{\mathrm{ex},2}/A_{\mathrm{ex},1}\). In this case,
\begin{equation}
	\frac{\omega_{M,2}\lambda_{2}^{2}}{\omega_{M,1}\lambda_{1}^{2}}
	=
	\frac{\beta_{A}}{\beta_{M}},
\end{equation}
and the symmetric-mode shift in Eq. (\ref{e:S-appr}) no longer vanishes when \(\beta_{A}\neq\beta_{M}\) and becomes $k$-dependent. The collective mode splitting, consequently, becomes $k$-dependent too. Such independent control of \(M_{\mathrm{s}}\) and \(A_{\mathrm{ex}}\) may thus offer an additional degree of freedom for optimizing the coupling length for a specific frequency range, or developing nonlinear devices, where $k$-dependence of the coupling length could be necessary \cite{wang2020magnonic}. 

\begin{figure}[!htp]
\centering
\includegraphics[width=0.8\textwidth]{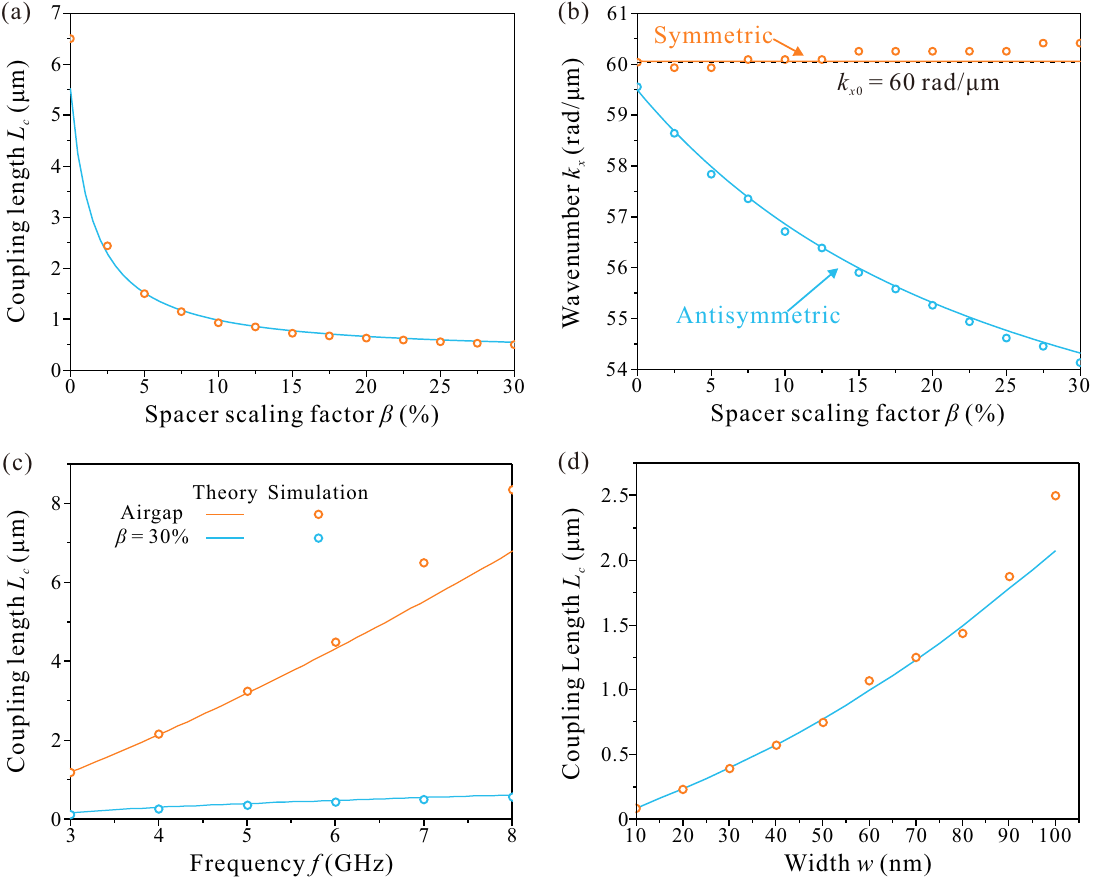}
\caption{\label{fig:cpl} \textbf{Dependence of the coupling characteristics on the magnetic spacer, operating frequency, and waveguide width.} The magnetic properties of the spacer are characterized by the scaling factor $\beta = M_{\mathrm{s},2}/M_{\mathrm{s},1} = A_{\mathrm{ex},2}/A_{\mathrm{ex},1}$, where $\beta = 0$ corresponds to an air gap. Unless varied explicitly, the operating frequency is 7 GHz. (a) Coupling length $L_{\mathrm{c}}$ as a function of $\beta$. (b) Wavenumbers of the symmetric and antisymmetric modes as functions of $\beta$. The horizontal dashed line denotes the reference wavenumber $k_{x0} = 60\ \mathrm{rad}/\mu\mathrm{m}$ corresponding to a single waveguide. (c) Coupling length as a function of frequency for the air-gap structure and a magnetic-spacer structure with $\beta = 30\%$. The coupling length of the air-gap structure increases rapidly with frequency, whereas the magnetic spacer substantially suppresses this increase. (d) Coupling length as a function of the individual waveguide width $w$ for $\beta = 15\%$. Solid lines show the analytical calculations, and open circles represent the micromagnetic simulation results. Unless varied explicitly, the waveguide width is 50 nm, the spacer width is 10 nm, and the waveguide and spacer thicknesses are 10 nm.}
\end{figure}

For the real system, both exchange and dipolar interactions contribute to the collective-mode dispersion. If the spacer is a weak magnetic material, its effect on the dynamic dipolar interaction can be treated as a small correction. The exchange and dipolar contributions then become simply additive:
\begin{equation}
	\omega_\t{\text{S},\text{AS}} = \omega_\t{i} \mp \frac{\Delta\omega_\t{dip}}2 + \Delta\omega_\t{\text{S},\text{AS}} ,
\end{equation}
where $\Delta \omega_{\mathrm{dip}}$ is the dipolar splitting between the collective modes. At small wavenumbers, the dipolar and exchange contributions coexist and both influence the mode separation. As $k_{x}$ increases, the dipolar contribution progressively decreases, whereas the exchange-induced shift of the antisymmetric mode remains finite. The magnetic spacer therefore prevents the modal splitting from collapsing when the spin waves enter the exchange-dominated regime, explaining the pronounced separation of the two branches in Fig.~\ref{fig:epsart}(b).

The analytical model consequently identifies two complementary roles of the magnetic spacer. It provides a continuous exchange pathway between the waveguides and introduces an additional exchange-energy cost for the antisymmetric collective mode. The same mode-selective frequency shift discussed above has also been observed for interlayer exchange coupling between magnetic layers: the antisymmetric mode shifts, whereas the symmetric mode remains unchanged \cite{barnas1989spin}. Together, these effects retain a substantial mode splitting at short wavelengths and reduce the corresponding coupling length.

The analytical model allows us to systematically determine how the magnetic properties of the spacer, the operating frequency, and the waveguide dimensions affect the exchange-mediated coupling. We characterize the spacer using a dimensionless scaling factor $\beta = \frac{M_{\mathrm{s},2}}{M_{\mathrm{s},1}} = \frac{A_{\mathrm{ex},2}}{A_{\mathrm{ex},1}}$. The two parameters are scaled by the same factor so that $A_{\mathrm{ex},2}/M_{\mathrm{s},2} = A_{\mathrm{ex},1}/M_{\mathrm{s},1}$. Here, $\beta = 0$ represents the air-gap structure, whereas $\beta > 0$ describes a magnetically ordered spacer. Unless otherwise specified, the results in Fig.~\ref{fig:cpl} are evaluated at a fixed frequency of 7 GHz.

Figure~\ref{fig:cpl}(a) shows the coupling length as a function of $\beta$. The coupling length decreases rapidly after a magnetic pathway is introduced between the waveguides. For the air-gap structure, the weak dipolar interaction in the exchange-dominated regime results in a coupling length of several micrometres. As $\beta$ increases, the exchange interaction across the spacer becomes progressively stronger, substantially increasing the modal splitting and shortening the coupling length. At $\beta = 30\%$, the coupling length is reduced to approximately 0.5 $\mu$m. Strong reduction occurs at relatively small values of \(\beta\), indicating that even a weakly magnetic spacer can produce a considerable enhancement of the coupling. The reduction in coupling length, however, gradually saturates with increasing \(\beta\), such that further increasing the magnetic properties of the spacer provides only a limited additional reduction in \(L_c\). In the limiting case of \(\beta=100\%\), the spacer becomes identical to the YIG waveguides, and the system effectively evolves into a single wider waveguide, where the power oscillation can instead be understood as beating between its transverse modes. Such a configuration, however, no longer provides two well-defined waveguides for efficient spin-wave injection and extraction, making it less suitable for realizing a directional coupler.

To reveal the modal origin of this reduction, Fig.~\ref{fig:cpl}(b) separately plots the wavenumbers of the symmetric and antisymmetric modes at 7 GHz. The horizontal dashed line indicates the reference wavenumber $k_{x0} = 60\ \mathrm{rad}/\mu\mathrm{m}$ corresponding to a single waveguide. As the magnetic properties of the spacer increase, the wavenumber of the symmetric mode remains close to the single waveguide value, whereas that of the antisymmetric mode decreases continuously. The enhancement of $\Delta k = |k_{\mathrm{s}} - k_{\mathrm{as}}|$ therefore originates predominantly from the shift of the antisymmetric branch.

This behavior follows directly from the theoretical model. Because $M_{\mathrm{s},2}$ and $A_{\mathrm{ex},2}$ are reduced by the same factor, $\omega_{\mathrm{M},2} \lambda_{2}^{2} = \omega_{\mathrm{M},1} \lambda_{1}^{2}$, and the exchange-induced shift of the symmetric mode vanishes within the analytical approximation. By contrast, the antisymmetric mode contains a strong transverse variation of the dynamic magnetization across the narrow spacer and therefore acquires an additional exchange-energy cost. At a fixed frequency, this upward frequency shift corresponds to a reduction in $k_{\mathrm{as}}$. Consequently, the magnetic spacer selectively shifts the antisymmetric mode while leaving the symmetric mode nearly unchanged, thereby producing the large wavenumber difference required for a short coupling length. It is worth noting that the nearly unchanged symmetric branch in Fig.~\ref{fig:cpl}(b) is a direct consequence of the proportional scaling \(\beta_{M}=\beta_{A}\); for a nonproportional modification, both branches shift.

Figure~\ref{fig:cpl}(c) directly compares the frequency dependence of the coupling length for the air-gap and magnetic-spacer structures. In the conventional air-gap coupler, $L_{\mathrm{c}}$ increases rapidly with frequency. 
The coupling length therefore reaches several micrometres in the exchange-dominated regime, despite the progressively shorter spin-wave wavelength.

Introducing a magnetic spacer strongly suppresses this increase. Although $L_{\mathrm{c}}$ still increases with frequency, its growth is much slower than that of the air-gap structure. 
Consequently, the magnetic-spacer coupler maintains a submicrometre coupling length over a broad frequency range and becomes increasingly advantageous as the operating frequency and wavenumber increase. The agreement between the analytical calculations and micromagnetic simulations confirms that the theoretical model captures this distinct frequency scaling.

Finally, Fig.~\ref{fig:cpl}(d) shows the effect of the individual waveguide width $w$ on the coupling length for a magnetic spacer with $\beta = 15\%$. The coupling length increases with increasing waveguide width. In narrower waveguides, the antisymmetric mode experiences a stronger transverse variation and hence a larger exchange-energy penalty, leading to a larger modal splitting and a shorter coupling length. This width dependence further demonstrates that exchange-mediated coupling becomes particularly effective as the transverse dimensions of the device are reduced. In all panels, the analytical calculations agree reasonably well with the micromagnetic simulations, validating the model over the investigated parameter range.

\begin{figure}[htb]
\centering
\includegraphics[width=0.8\textwidth]{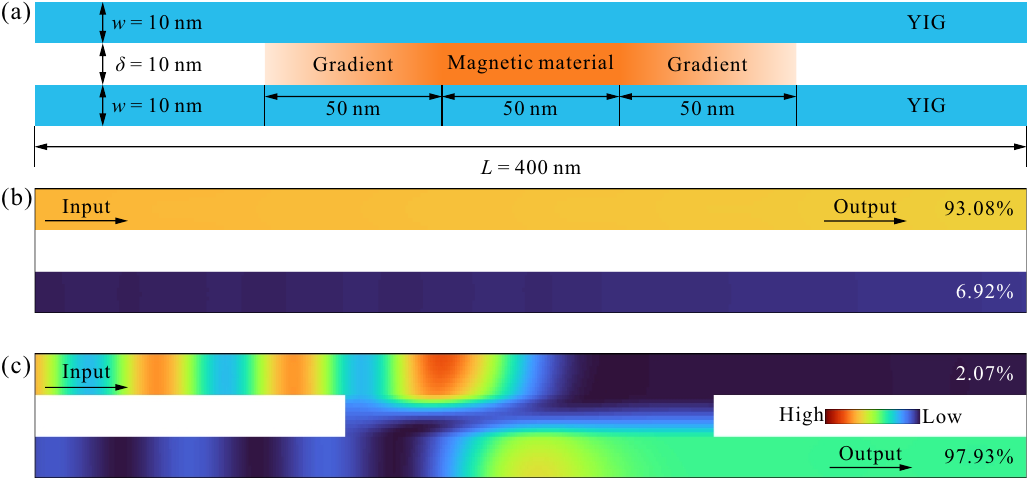}
\caption{\label{fig:DC}\textbf{Ultracompact directional coupling of short-wavelength spin waves.} (a) Schematic of the straight two-waveguide device. Each YIG waveguide is 10 nm wide, and the air gap between them is 10 nm wide, giving a total lateral width of 30 nm. The device is 400 nm long. Within the 150-nm-long coupling region, the spacer scaling factor $\beta$ increases linearly from 0 to 15\% over the first 50 nm, remains at 15\% over the central 50 nm, and decreases linearly to 0 over the final 50 nm. (b, c) Micromagnetically simulated spin-wave intensity distributions at 7 GHz (b) without and (c) with the magnetic spacer. Without the spacer, 93.08\% of the normalized output power remains in the input waveguide and only 6.92\% reaches the adjacent waveguide. With the graded magnetic spacer, 97.93\% of the normalized output power is transferred to the adjacent waveguide, while 2.07\% remains in the input waveguide.}
\end{figure}

\raggedbottom
Having established the physical origin and parameter dependence of the exchange-mediated coupling, we next demonstrate how it enables an ultracompact directional coupler for short-wavelength spin waves. Previous dipolar spin-wave directional couplers generally employ curved access waveguides to increase their separation outside the coupling region \cite{wang2018reconfigurable,ge2025deeply,szulc2025magnetic,ren2019reconfigurable,tian2025non}. The increased separation suppresses unwanted dipolar coupling and allows the spin waves to be injected into and extracted from well-defined individual waveguides. This approach, however, considerably enlarges the lateral footprint of the device.

For the ultrashort spin waves considered here, the dipolar interaction between two waveguides separated by an air gap is already extremely weak. The parallel waveguides are therefore naturally decoupled outside the active coupling region, eliminating the need for curved input and output sections. As illustrated in Fig.~\ref{fig:DC}(a), we use two straight YIG waveguides, each 10 nm wide, separated by a 10-nm-wide gap. The resulting device has a total lateral width of only 30 nm and an overall length of 400 nm.

Exchange-mediated coupling is activated locally by introducing a 150-nm-long magnetic spacer between the waveguides. To mitigate spin-wave reflection caused by an abrupt change in the local dispersion at the boundaries of the coupling region, the magnetic parameters of the spacer are varied gradually along the propagation direction. The scaling factor $\beta$ increases linearly from 0 to 15\% over the first 50 nm, remains constant at 15\% over the central 50 nm, and then decreases linearly to 0 over the final 50 nm. These graded transition regions provide a smooth conversion between the uncoupled modes of the air-gap waveguides and the collective modes of the exchange-coupled section. Such a graded magnetic profile is also experimentally feasible and could be realized by spatially controlling the irradiation dose, for example, through the local exposure time of a focused laser or ion beam\cite{florio2026programmableintegratedmagnonicmeshes,kiechle2023spin,naunheimer2026establishing}.

Figure \ref{fig:DC}(b) and \ref{fig:DC}(c) compare the simulated spin-wave propagation at 7 GHz without and with the magnetic spacer, respectively. In the air-gap structure, the two straight waveguides remain almost decoupled: 93.08\% of the normalized output power remains in the excited waveguide, while only 6.92\% reaches the adjacent waveguide. When the graded magnetic spacer is introduced, the enhanced splitting between the symmetric and antisymmetric modes drives coherent power transfer across the 150-nm-long coupling region. At the output, 97.93\% of the normalized power is routed to the adjacent waveguide, with only 2.07\% remaining in the input waveguide. The exchange-mediated structure thus reverses the output distribution and achieves approximately 98\% transfer without curved waveguides.

The complete coupler occupies an area of only 400 nm $\times$ 30 nm, demonstrating that exchange-mediated directional coupling can translate the short wavelength of exchange spin waves into a comparably small device footprint. Further optimization of the spacer profile, coupling-region length, and waveguide dimensions may reduce the footprint and residual power in the undesired output even further. 

In practice, however, the laser or ion-beam modification used to define the spacer may also substantially increase its magnetic damping\cite{giacco2024patterning,bensmann2025dispersion}. It is therefore essential to determine whether this fabrication-induced loss compromises the coupling performance.

\begin{figure}[!htp]
\centering
\includegraphics[width=0.8\textwidth]{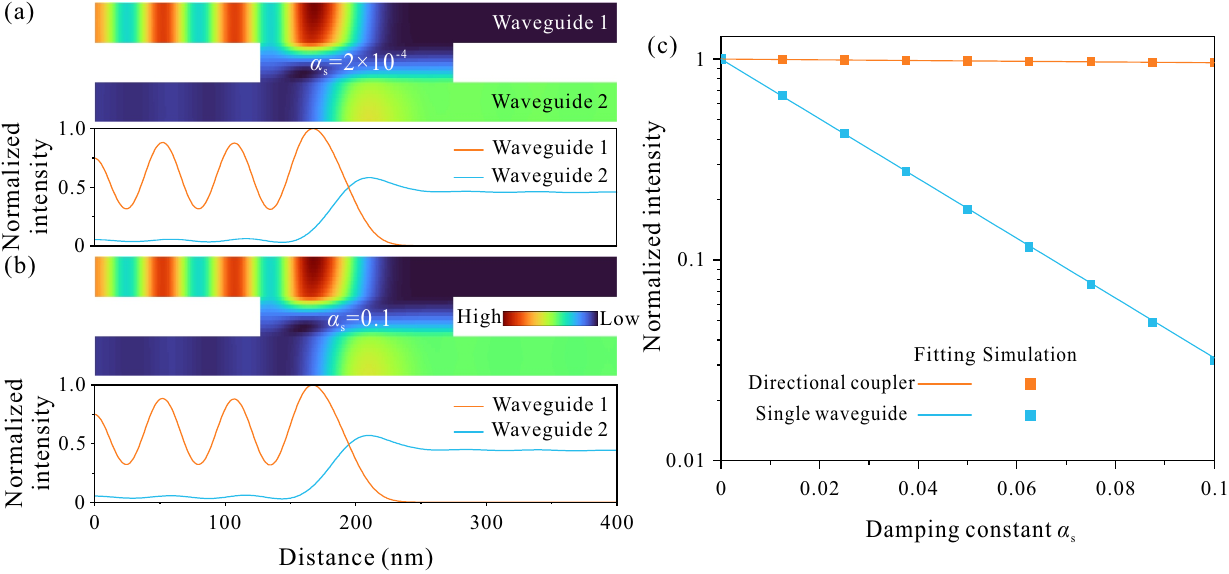}
\caption{\label{fig:damping} \textbf{Robustness of the exchange-mediated directional coupler against damping in the magnetic spacer.} (a,b) Micromagnetically simulated spin-wave intensity distributions and the corresponding longitudinal intensity profiles in waveguides~1 and~2 for spacer damping coefficients of (a) \(\alpha_{\mathrm{s}}=2\times10^{-4}\) and (b) \(\alpha_{\mathrm{s}}=0.1\). (c) Normalized spin-wave intensity extracted at the output of waveguide~2 as a function of the damping coefficient. The orange data show the effect of varying the spacer damping while keeping the waveguide damping fixed at \(2\times10^{-4}\). The blue data show the effect of varying the damping of the propagating YIG waveguide. Symbols represent micromagnetic simulations, and solid lines show the corresponding fits.}
\end{figure}

To evaluate this fabrication-related constraint, we compare the operation of the coupler over a wide range of spacer damping. Fig.~\ref{fig:damping}(a) shows the simulated spin-wave intensity distribution and the corresponding intensity profiles in the two waveguides for a low spacer damping of \(\alpha_{\mathrm{s}} = 2 \times 10^{-4}\). The oscillations in waveguide 1 originate from interference between the incident and partially reflected spin waves. This indicates that the graded spacer substantially reduces reflection at the coupling-region boundaries, although it does not completely eliminate it.

Figure~\ref{fig:damping}(b) shows the corresponding result when the spacer damping is increased by nearly three orders of magnitude to \(\alpha_{\mathrm{s}}=0.1\). Remarkably, both the spatial distribution and the transmitted spin-wave intensity remain almost unchanged. This robustness arises because the propagating spin-wave power is predominantly confined to the two YIG waveguides, whereas the spacer mainly provides the transverse exchange interaction required to hybridize their collective modes. Consequently, only a small fraction of the total mode energy is dissipated in the spacer, even when its local damping is large. This loss-avoiding behavior is reminiscent of the magnonic stimulated-Raman adiabatic-passage mechanism, in which efficient transfer can be maintained while the lossy intermediate state remains only weakly populated \cite{wang2021stimulated,ivanov2005spontaneous}.

The different roles of spacer and waveguide damping are summarized in Fig.~\ref{fig:damping}(c). Increasing \(\alpha_{\mathrm{s}}\) has only a minor influence on the spin-wave intensity measured at the output of waveguide 2. By contrast, increasing the damping of the propagating waveguide causes a strong exponential reduction of the output intensity because the spin wave continuously accumulates propagation loss. The micromagnetic results are well described by an exponential-decay fit. These results demonstrate that the exchange-mediated coupler remains functional even when the magnetic modification used to create the spacer introduces substantial local damping, which considerably relaxes the material requirements for its experimental realization.

In conclusion, we have proposed an exchange-mediated directional coupler that overcomes the short-wavelength limitation of conventional dipolar spin-wave couplers. By replacing the nonmagnetic gap between two YIG waveguides with a weak magnetic spacer, a continuous exchange pathway is established, preserving strong coupling in the exchange-dominated regime. The magnetic spacer reduces the coupling length by approximately 13 times compared with the conventional air-gap structure.

The analytical model shows that the enhanced coupling originates predominantly from the exchange-induced shift of the antisymmetric mode, while the symmetric mode remains nearly unchanged. Based on this mechanism, we designed a straight directional coupler with an area of only \(0.012~\mu\mathrm{m}^2\), which transfers approximately \(98\%\) of the normalized output power to the target waveguide without requiring curved input and output sections. The coupling performance is also insensitive to a strong increase in spacer damping because the propagating spin-wave power remains predominantly confined to the low-damping waveguides. These results provide a route toward experimentally feasible, ultracompact routing elements and highly integrated magnonic circuits operating with short-wavelength exchange spin waves.


%
%

%


\section*{Methods}
Micromagnetic simulations were performed using MuMax3\cite{vansteenkiste2014design}. Unless otherwise specified, each waveguide had a width of \(w=50~\mathrm{nm}\), a thickness of \(t=10~\mathrm{nm}\), and a length of \(L_{\mathrm{w}}=1000~\mathrm{nm}\). The separation between the two waveguides was \(d=10~\mathrm{nm}\). The simulation region was discretized into cells of \(1~\mathrm{nm}\times1~\mathrm{nm}\times10~\mathrm{nm}\), corresponding to one cell along the film thickness. Regions with gradually increasing damping were introduced at both ends of the waveguides to suppress spin-wave reflections. A spatially uniform sinc-shaped magnetic field was applied within a 5 nm range at one end of the waveguides to excite spin waves over a broad frequency range.

To obtain the spin-wave dispersion, the spatial and temporal evolution of the dynamic magnetization \(m(x,y,t)\) was recorded. For each transverse position \(y\), a two-dimensional fast Fourier transform was performed with respect to the propagation coordinate \(x\) and time \(t\), yielding the wavevector-frequency spectrum \(S(k_x,\omega;y)\). The spectral intensity was then averaged over the transverse direction to obtain the dispersion relation \(S(k_x,\omega)\). The wavenumbers of the symmetric and antisymmetric modes were extracted from the corresponding spectral maxima at each frequency.

\begin{acknowledgments}
 This work was supported by the National Natural Science Foundation of China (Grant No. 12574118). R. V. acknowledges support by the National Research Foundation of Ukraine (Grant No. 2025.07/0132). P. P. acknowledges support by the Deutsche Forschungsgemeinschaft (DFG, German Research Foundation) -TRR 173--268565370 (“Spin + X”, Project B01) and by the European Research Council within the Starting Grant No. 101042439 “CoSpiN”. A. V. C. acknowledges the financial support of the Austrian Science Fund (FWF) by means of grant MagNeuro no. 10.55776/PIN1434524. 
\end{acknowledgments}

\section*{Author Contributions}
Z. Z. performed micromagnetic simulations. R. V. derived the theoretical model. A. V. C., P. P., and Q. W. led this project. Q. W. conceived the idea. Z. Z. and Q. W. wrote the manuscript with the help of all the coauthors. All authors contributed to the scientific discussion and commented on the manuscript.

\bibliography{myref}

\end{document}